\documentclass{elsart}

\usepackage{graphicx}

\usepackage{amssymb}

\begin{document}

\begin{frontmatter}


\title{Exploring QCD: from LEAR to GSI}
\author{T.Barnes}
\address{Physics Division, Oak Ridge National Laboratory\\
and\\
Department of Physics and Astronomy, University of Tennessee
}

\title{}


\author{}

\address{}

\begin{abstract}
In this invited contribution I briefly review some of the principal 
topics in hadron spectroscopy that were studied at the CERN 
low-energy antiproton facility LEAR, from its beginnings in the early
1980s to the present. These topics include the nature of
multiquark systems, the short-ranged nuclear force, and gluonic hadrons,
including glueballs and hybrids. Lessons we have learned from the LEAR 
program that are relevant
to the future GSI project are given particular emphasis.
\end{abstract}

\end{frontmatter}


\section{Introduction}
\label{sec1}
The title suggested for this invited talk included 
the phrase ``Testing QCD". To me this implies a rather better 
theoretical understanding of low-energy QCD spectroscopy than
we have at present. As this largely historical
overview will show, experimental discoveries
in this field have repeatedly surprised theorists, 
and the resonances
found often bear only a passing resemblance
to our predictions. In light hadron spectroscopy, 
we are still exploring unknown territory.

The fact that we have so much to learn from 
experiment
is one of the great attractions
of this field. 
Although the starting point of the QCD lagrangian
is universally accepted, even our best theoretical methods
(specifically lattice gauge theory) currently involve 
uncontrolled approximations, and the resulting predictions need not
closely resemble what is found experimentally in the real hadron spectrum. 
Progress in lattice gauge theory and in the development of
models of hadrons continues to benefit from close comparisons to 
experiment.

Much of the recent experimental progress in hadron spectroscopy has come from
nucleon-antiproton collisions, especially at LEAR. The parallel experimental
efforts using meson beams, $e^+e^-$ and other initial states have as always
been valuable as independent confirmations of discoveries, or for suggestions
regarding interesting final states for detailed studies. In this review I will
sketch the history of our ideas regarding hadron spectroscopy, first as it was
understood at the beginning of the LEAR program, then how it developed and 
was modified by results from LEAR and elsewhere, and finally where we 
stand today. I will conclude with a few suggestions for interesting topics that
might be addressed at the future GSI antiproton facility.
   
\section{Classification of hadrons}
\label{sec2}

Our general classification of hadrons as color-singlet states of quarks
and gluons has not changed since the early days of QCD. The simplest 
such states are the conventional quark model $qqq$ baryons and $q\bar q$ 
mesons, and we have {\it ca.} 10$^2$ examples of experimental resonances that
appear to be well-described as such states. 

In addition to these simplest possible color singlets, other combinations of
quarks, antiquarks and gluons also span color singlets, and might in 
principle be expected to appear as resonances in the experimental 
light hadron spectrum. These include the $q^2\bar q^2$ multiquark states
known as ``baryonia", pure glue basis states such as $gg$ and $ggg$
``glueballs", and mixed quark-and-gluon states known as ``hybrids", which
include $q\bar q g$ hybrid mesons and $qqqg$ hybrid baryons. Of course these
simple basis states will mix when the quantum numbers allow, so that we expect
the physical resonances to be strongly mixed linear combinations of such
basis states. The level of configuration mixing 
and the types of basis state that best describe the physical resonances 
are at present open and poorly understood issues.

\section{LEAR hadron physics goals}
\label{sec3}

\subsection{Baryonia: LEAR ab initio}
\label{sec3a}

The principal initial motivation for LEAR was to search for the novel
``baryonium" $q^2\bar q^2$ mesons. These states had been studied in
many models, such as the MIT bag model, quark potential models, color
chemistry models, and so forth, and a very rich spectrum of discrete
levels was predicted. 

The purported connection of $q^2\bar q^2$ baryonium states to
$p\bar p$ annihilation is quite interesting, because it introduces a
crucial and still poorly understood topic, which is the nature of the 
short-ranged nuclear force. Progress in understanding this force 
should be a major goal of future studies in antiproton physics, for example
at GSI. The argument leading to a strong coupling between 
$p\bar p$ annihilation and $q^2\bar q^2$ resonances was the belief that
the short-ranged nucleon-nucleon repulsive core was due primarily to 
vector meson $(\omega)$ exchange. This was discussed for example 
by Richard 
\cite{Ric82} 
at the 1982 Erice LEAR meeting.
If this description is accurate, one can presumably similarly describe
the corresponding $p\bar p$ $\omega$-exchange force after a G-parity 
transformation, and an attractive core is found which supports deeply 
bound $p\bar p$ states. (Hence the name ``baryonia".) The connection to
$q^2\bar q^2$ requires a bit more imagination. If one takes a quark line
diagram for $p\bar p \to p\bar p$ with $q\bar q$ meson exchange in 
t-channel, and deforms the exchange so the $q$ and $\bar q$ lines are
widely separated, one sees intermediate $q^2\bar q^2$ states in s-channel.
Thus one can argue that if the short-ranged nuclear force is indeed due to
t-channel vector meson exchange, there should be deeply bound $p\bar p$
states with a strong coupling to $q^2\bar q^2$. Since there were many models
that suggested a rich spectrum of discrete $q^2\bar q^2$ levels in the 1970s,
a rather undermotivated
identification of bound $p\bar p$ states with $q^2\bar q^2$ was suggested.

Even more remarkably, it was widely suggested (with little support from theory)
that these light $q^2\bar q^2$ resonances might be very narrow ($\Gamma_{tot}$ = few MeV),
and that these narrow states might be evident for
example in the total $p\bar p$ cross sections. Ambiguous evidence for narrow
baryonium candidates such as the ``S(1936)" was frequently cited in the literature
of the early 1980s \cite{Wal82}.
One of the clear experimental conclusions of the subsequent LEAR program was the 
absence of such narrow states \cite{Bra84}.   

Unfortunately the confining interaction required
to give this discrete $q^2\bar q^2$ spectrum was a model artifact;
$q^2\bar q^2$ states can in practice simply ``fall apart" into two
separate $q\bar q$ mesons when energetically allowed, 
and hence need not exist as resonances. 
This erroneous prediction of a rich spectrum of light multiquark resonances
was referred to by Isgur as the ``multiquark fiasco" \cite{Isg85}.  

We note in passing that for sufficiently large quark mass 
``heavy-light" $Q^2\bar q^2$ bound states do lie below these 
fall-apart thresholds, 
and hence should exist as resonances \cite{Ric00}. 
However it is unclear whether $c$ quarks are sufficiently heavy to form
such
bound states; one might instead require a much more 
difficult study of the $b^2 \bar q^2$ system.

The fact that predictions of light $q^2\bar q^2$ states relied on a 
meson-exchange model of short-ranged nuclear forces highlights the 
importance of understanding the physical mechanisms underlying
these forces. An alternative description at the quark-gluon level, which
has been confirmed by many independent theoretical calculations, finds
instead that the nucleon-nucleon repulsive core is well described by the
one-gluon exchange spin-spin contact interaction. If this picture is
correct, there is no simple relation between the $pp$ and $p\bar p$
interactions, and there need not be deeply-bound $p\bar p$ bound states.
Evidently these different models of interhadron forces lead to very different
predictions of the bound state spectrum. It is clearly of great importance to
determine which mechanism, if either, is dominant at short distances.
Studies of the reaction $p\bar p \to \Lambda \bar \Lambda$ at LEAR 
have attempted to clarify this issue, and the most recent results 
were presented at this meeting. Unfortunately the conclusions are
rather ambiguous at present; neither meson exchange nor quark-gluon models
in their present incarnations give 
a good description of the data 
\cite{Kin03}. This may not be surprising, since one would expect meson-exchange
(e.g. kaon-exchange) to correctly describe the long-distance part 
of the interaction, but this picture should fail at sufficiently short distances 
where the baryon and antibaryon three-quark and three-antiquark
wavefunctions experience considerable overlap.

\subsection{Glueballs}
\label{sec3b}

\begin{figure}[ht]
\begin{center}
\includegraphics[width=0.8\textwidth,angle=0]{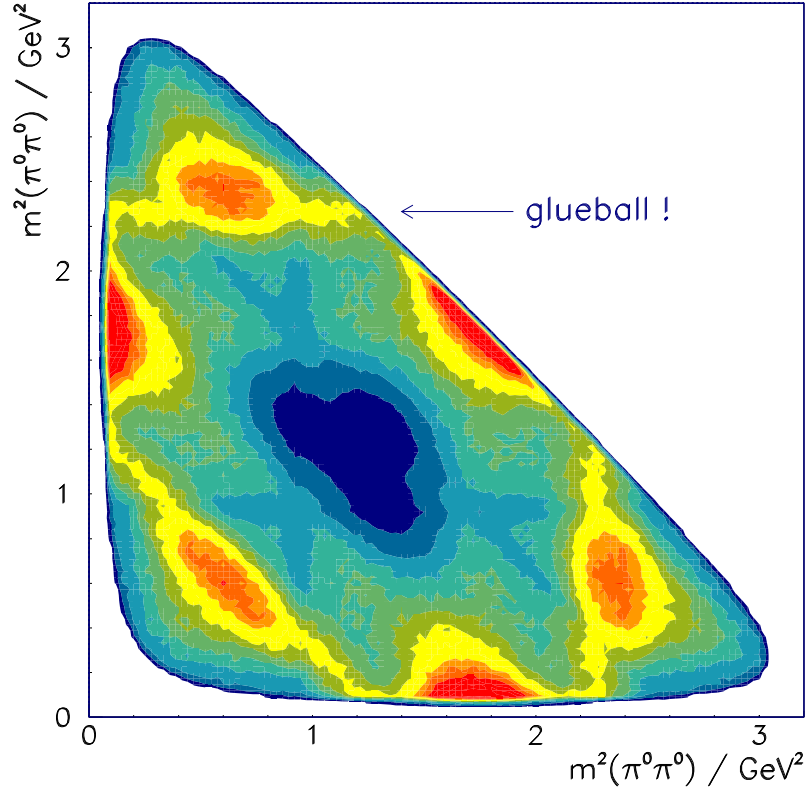}
\vskip -3cm
\caption{Dalitz plot of $p\bar p \to \pi^o \pi^o \pi^o $ showing the
$f_0(1500)$ glueball candidate \cite{Ams95}.
}
\label{fig:3pi0col}
\end{center}
\end{figure}

Not long after the beginning of planning for the LEAR program, much
interest arose in the possibility of observing gluonic excitations
in the meson spectrum \cite{Bar82}, in addition to the multiquark states that were
the original motivation for LEAR. 
Theorists noted that color-singlet states could be formed in the pure
glue sector from $|gg\rangle, |ggg\rangle \dots$ states, and in addition
mixed quark-and-gluon color-singlet basis states such as 
$|q\bar q g\rangle$ were allowed. 
The associated resonances are now known
as glueballs and hybrids respectively.
Detailed predictions for the spectra of these states were published,
most notably using the MIT bag model \cite{bag_glue}
and lattice gauge theory (LGT)
\cite{Hal83}.

Interest in glueballs at LEAR also arose from 
the experimental $J/\psi$ radiative
decay program, which had identified the glueball candidates 
``$\iota$"
(now the $\eta(1440)$, and excluded as a glueball due to its low mass) and the 
``$\theta$", initially thought to be 
$2^{++}$ but now known as the $f_0(1710)$, and widely considered 
a scalar glueball candidate. It is amusing that LGT calculations at
that time claimed $0^{-+}$ and $2^{++}$ glueballs at masses
of {\it ca.} 1420 and 1620~MeV respectively, 
consistent with these newly discovered states. 

One might expect to find the most striking evidence of new states 
in the glueball sector, since these are the farthest removed physically from
conventional $q\bar q$ states. In practice this does not appear to be the case.
The theoretical ``pure glue" glueball spectrum is now quite well understood 
from LGT studies \cite{Mor99}, 
and is rather sparse at the energies
of relevance to LEAR; the only glueball predicted to have a mass below 2~GeV
is a scalar, at a mass of about 1.7~GeV. The next glueballs 
predicted with increasing mass are  
$0^{-+}$ and $2^{++}$ states near 2.5~GeV; the first exotic encountered
in the LGT glueball spectrum is a $2^{+-}$ state, at a very high mass
(near 4~GeV). As experimental studies of the light meson spectrum
are almost exclusively concerned with the region below 2.5~GeV, the
subject of glueballs specializes to the search for a single I=0 scalar state.
Unfortunately, this sector is one of the most obscure in the light meson
spectrum.

There are currently two candidates for the scalar glueball, the $f_0(1500)$ from
LEAR and the $f_0(1710)$, originally reported in $J/\psi$ radiative decays.

The $f_0(1500)$ 
is clearly visible in the beautiful high-statistics 
$p\bar p \to 3\pi^o$ Dalitz plot shown in Fig.1, 
made famous by the Crystal Barrel
Collaboration \cite{Ams95}. This state may be the single most interesting 
discovery of the LEAR hadron physics program. As an ``extra" I=0
scalar state near 1.7~GeV it is a natural glueball candidate,
and the width of $\approx 100$~MeV appears anomalously small for a light
$^3$P$_0$ $q\bar q$ quark model state. 

There are problems however with
identifying either the $f_0(1500)$ or the $J/\psi$-radiative state
$f_0(1710)$ with the scalar glueball. Naively one would expect approximate
SU(3) flavor symmetry in the decay couplings of a glueball, which should
lead to relative PsPs couplings of 
$\pi\pi: KK: \eta\eta: \eta\eta': \eta'\eta' =
3:4:1:0:1$. (These are branching fractions divided by phase space.)
Since the observed branching fraction ratios are
$B_{KK/\pi\pi}(f_0(1500)) = 0.19\pm 0.07$   
and
$B_{\pi\pi/KK}(f_0(1710)) = 0.39\pm 0.14$, 
neither state decays with the nearly equal
$\pi\pi$ and $KK$ branching fractions expected for a glueball.
This problem may be due to mixing between $q\bar q$ and $G$ basis states
\cite{Ams96},
or it may imply that there is strong momentum dependence in these decay 
couplings.  

\vskip 1cm
\subsection{Hybrids}
\label{sec3c}

Shortly after interest arose in the glueball spectrum it was realized
that a rather richer spectrum of $q\bar q g$ ``hybrid meson" states should
also exist. There are many more experimental opportunities to search for hybrids
than glueballs because (light) hybrids span flavor nonets, and it
was also found that the lightest hybrid multiplet includes 
very characteristic $J^{PC} = 1^{-+}$ exotic quantum numbers \cite{hybr_bag}.
The bag model estimates for the mass of this light hybrid exotic were typically about
1.5~GeV. This work was followed by studies using 
the flux-tube model \cite{hybr_ft} and LGT \cite{hybr_lgt}, which find a rather
higher mass of about 1.9-2.1~GeV for this $1^{-+}$ exotic.

\begin{figure}[ht]
\begin{center}
\includegraphics[width=1.0\textwidth,angle=0]{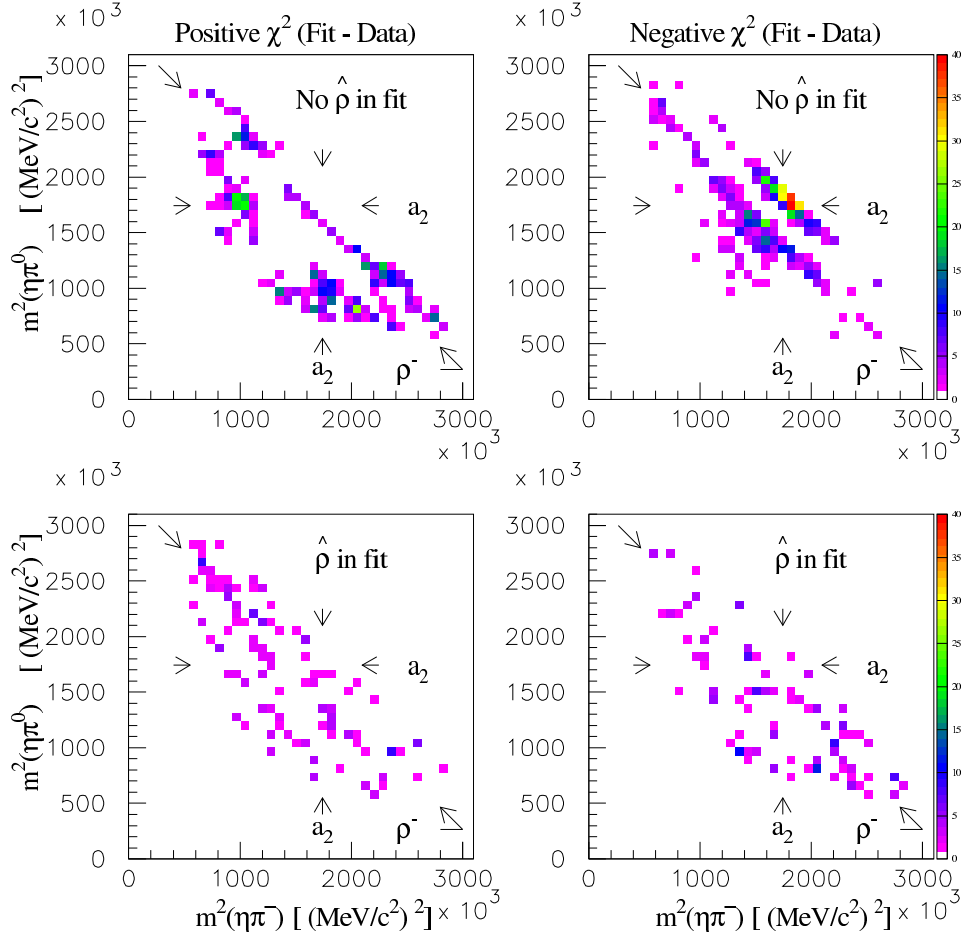}
\vskip -5cm
\caption{A difference Dalitz plot of 
$n\bar p \to \pi^+ \pi^o \eta $ showing structures (upper panels)
that are accounted for by assuming a broad $\pi_1(1400)$ exotic resonance
(lower panels) \cite{pi1_1400_CB}.
}
\label{fig:pi1_1400}
\end{center}
\end{figure}

\vskip 0.3cm
The flux-tube model has also been applied to 
strong decays of hybrids
\cite{hybr_decays}, which leads to the well-known flux-tube selection
rule that light hybrids typically decay preferentially into S+P final states,
in which one final $q\bar q$ meson has an orbital excitation. (Note however
that the present exotic hybrid candidates do not support this rule.)  

\vskip 0.3cm
Experimentally there has been great interest in searching for the very 
characteristic $J^{PC} = 1^{-+}$ exotics, especially with I=1 flavor. Two such
states have been reported, a $\pi_1(1400)$ and a $\pi_1(1600)$. Note that
both candidates are rather lower in mass than the LGT and flux-tube expectations
of $M\approx 1.9$-$2.1$~GeV.
The $\pi_1(1400)$ (first reported by VES and BNL in 
$\eta\pi$ 
\cite{pi1_1400} after a long history of studies of this channel) 
is the more controversial, since this is a rather weak, broad
signal in a channel that is dominated by the $a_2(1320)$.  
This state has apparently been confirmed in $p\bar p$ annihilation
at LEAR by the Crystal Barrel Collaboration \cite{pi1_1400_CB}; it is notable
that the coupling of $p\bar p$ to this candidate exotic is quite large, 
which argues in favor of $p\bar p$ facilities as an approach for the study of
exotic mesons. The second $J^{PC} = 1^{-+}$ exotic candidate is the
$\pi_1(1600)$, also reported by VES and E852 \cite{pi1_1600}, in the channels
$b_1\pi$, $\eta \pi$ and $\eta' \pi$. 
The $\pi_1(1600)$ appears remarkably clearly 
in the recent analysis of the $\eta'\pi^-$ channel by E852  
\cite{pi1_1600_pietap}, due largely to the weak 
coupling of $\eta' \pi$ to other resonances in this mass region.
(Fig.2 of this reference is our Fig.3; the $\pi_1(1600)$ is in panel $c$.) 
This very clear evidence for
a $J^{PC}$-exotic is strong motivation for future searches for the 
flavor partners of this state, as well as the other $J^{PC}$ states
expected near mass if this is indeed a light hybrid. The flux-tube model
anticipates 72 approximately degenerate states in the lightest hybrid multiplet,
and many of these should be narrow enough to be observable.
If the partner states are not observed, the 
$\pi_1(1600)$ would appear doubtful as a hybrid candidate.

\begin{figure}[ht]
\begin{center}
\includegraphics[width=0.8\textwidth,angle=0]{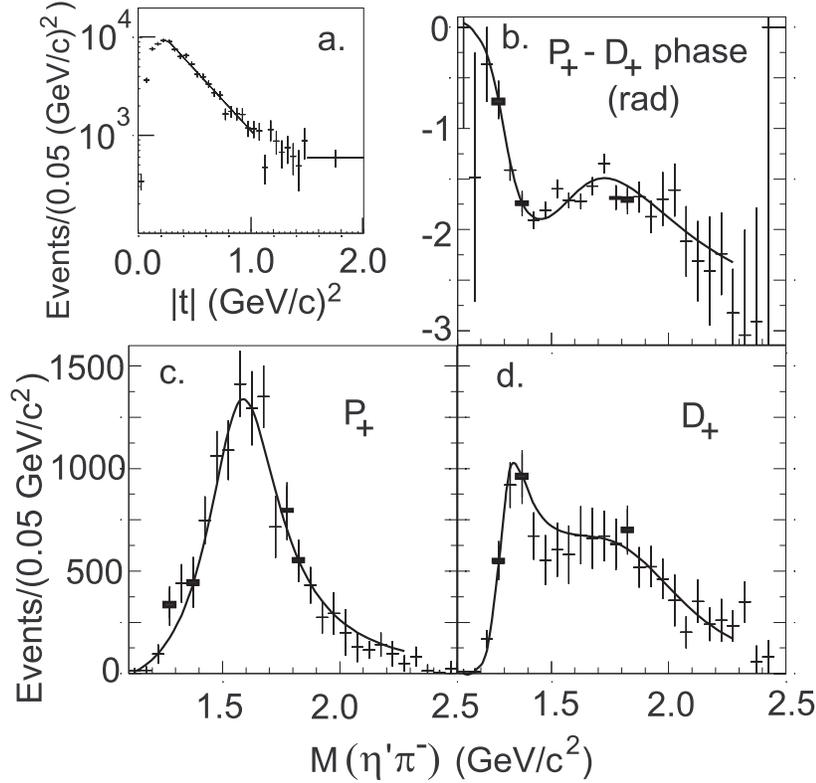}
\vskip -0cm
\caption{The $\pi_1(1600)$ reported in
$\pi^- p \to \eta' \pi^- p$ by the E852 Collaboration \cite{pi1_1600_pietap}.
The exotic $1^{-+}$ wave is shown in the bottom left panel.
}
\label{fig:pi1_1600}
\end{center}
\end{figure}

\vskip 1cm
\subsection{Quarkonia}
\label{sec3d}

Although the study of conventional $q\bar q$ mesons does not
carry the same excitement as the discovery of new types of states,
the identification of 
higher-mass $q\bar q$ mesons ($q=u,d,s$) has 
nonetheless been one of the most important 
experimental activities at LEAR and other hadron machines over the previous
two decades. The 
identification of any anomalous states will be much easier
if we have a reasonably complete description of the background of ``ordinary"
$q\bar q$ mesons. Since non-$q\bar q$ mesons are expected to 
first appear in the mass range 1.5-2.0 GeV, identification of the 
conventional quark model states over this mass range is 
a correspondingly important exercise.

Many of these $q\bar q$ states were unknown or poorly established until the last decade,
and much progress has been made recently. For example, of the 44 
$n\bar n$ mesons ($n=u,d$) predicted to lie below 2.1 GeV 
(spanning the 1S, 2S, 3S, 1P, 2P, 1D and 1F $n\bar n$ multiplets), 
about 35 have now been reported. Recent discoveries include 
candidates for the 2P and 1F multiplets that lie
in the crucial 1.5-2.0 GeV mass range.  
These multiplets are now experimentally about 80\% complete, which is 
comparable to how well we know the excited kaon spectrum in this mass range
({\it ca.} 16 of 22 states known). In comparison the $s\bar s$ spectrum is a
{\it terra incognita}, with only about 6 of these 22 states well established.

We do note that there is evidence of disagreement between theory and
experiment in the light quarkonium spectrum, notably in radial excitations.
One important discrepancy is in the prediction of the 
energy scale of radial excitations of $q\bar q$ states. For example,
Godfrey and Isgur \cite{God85} predicted that the 2P excitation of the 
light $h_1(1170)$ should lie at 1.78~GeV, whereas the candidate reported
recently by E852 is at 1.59~GeV \cite{h1expt}.
The reports of possible hybrids near 1.4 and 1.6 GeV suggest that we may
find a rich overpopulation of the meson spectrum relative to the 
predictions of the $q\bar q$ quark model beginning at about this mass.
High-statistics partial wave analyses 
will be important for the identification of conventional $q\bar q$ mesons
as well as the non-exotic hybrids, glueballs, multiquark systems, and 
mixing effects between these states 
that may be evident in this mass range.

\section{From LEAR to GSI}
\label{sec4}

To the extent that 
the intended purpose of LEAR was to identify new types of hadrons, it has been
successful. We now have two glueball candidates,
the $f_0(1500)$ (from LEAR) and the $f_0(1710)$, and two exotics, the
$\pi_1(1400)$ (confirmed by LEAR) and the $\pi_1(1600)$. There 
is no clear evidence for multiquarks as originally proposed. 
We note in passing however that the
$f_0(980)$ and $a_0(980)$ states were clearly observed at LEAR, for example
in $p\bar p \to \eta \pi \pi$, and are widely believed
to have large multiquark components. The closely related and crucially important 
subject of the nature of short-ranged nuclear forces remains 
under investigation, for example using 
LEAR data on $p\bar p \to \Lambda \bar \Lambda$. 

Regarding glueballs and hybrids, the existing experimental candidates 
represent a rather thin and ambiguous 
beginning to what should eventually 
become a rich field of spectroscopy. Neither 
glueball candidate decays according to expectations for a flavor singlet, 
which suggests that 
mixing between glueball and $q\bar q$ basis states may be very important. 
Of the 72 resonances expected in the lightest
hybrid multiplet near 1.9~GeV, 
we have just two candidates for 1~exotic hybrid 
(the $\pi_1$ level), and in 
neither case do the reported decay modes agree with theoretical
expectations for hybrids.

In conclusion, what we have from LEAR is ``proof of principle". 
An overpopulation of light scalars exists, giving us glueball and
multiquark/molecule 
candidates, 
and $J^{PC}$-exotic mesons exist, giving us exotic hybrid candidates. 
However there is thus far little evidence for 
the partner states that should also be present if these are indeed glueball
and hybrid states. A primary task for GSI will be to confirm
or refute the existing gluonic candidates, 
establish their strong branching fractions and other characteristic properties,
and search for multiplet partner states with other flavors (u,d,s,c) \cite{hybr_cc}
and $J^{PC}$ quantum numbers that must also
exist if LEAR's discoveries are indeed gluonic hadrons.

\end{document}